\documentclass[preprint,5p,twocolumn]{elsarticle}

\usepackage{graphicx}
\usepackage{amssymb}
\usepackage{subfigure}
\usepackage{amsmath}
\DeclareMathOperator*{\argmin}{argmin} 
\usepackage{amssymb}
\usepackage{listings}
\usepackage{graphicx}
\usepackage[rightcaption]{sidecap}
\usepackage{textcomp}
\usepackage{array} 
\usepackage{booktabs} 
\usepackage{multirow} 
\usepackage{caption} 
\usepackage{floatflt}
\usepackage{subfigure}
\usepackage{rotating} 
\usepackage{scrextend}
\usepackage[ruled,vlined]{algorithm2e}
\usepackage{listings}
\usepackage{hyperref}

\setcitestyle{authoryear,open={(},close={)}}

\journal{Journal Name}

\begin{document}

\begin{frontmatter}


\title{Dynamical strategies for obstacle avoidance during \textit{Dictyostelium discoideum} aggregation: a Multi-agent system model.}



\author[label1,label2]{Daniele Proverbio\corref{cor1}}
\ead{john.smith@uni.lu}
\author[label1,label3]{Marco Maggiora}

\address[label1]{University of Turin, Department of Physics (IT)}
\address[label2]{University of Luxembourg, Luxembourg Centre for Systems Biomedicine (LU)}
\address[label3]{Section of Turin, INFN, IT}

\cortext[cor1]{Correspondence:}

\begin{abstract}
Chemotaxis, the movement of an organism in response to chemical stimuli, is a typical feature of many microbiological systems. In particular, the social amoeba \textit{Disctyostelium discoideum} is widely used as a model organism, but it is not still clear how it behaves in heterogeneous environments. A few models focusing on mechanical features have already addressed the question; however, we suggest that phenomenological models focusing on the population dynamics may provide new meaningful data. Consequently, by means of a specific Multi-agent system model, we study the dynamical features emerging from complex social interactions among individuals belonging to amoeba colonies.\\
After defining an appropriate metric to quantitatively estimate the gathering process, we find that: a) obstacles play the role of local topological perturbation, as they alter the flux of chemical signals; b) physical obstacles (blocking the cellular motion and the chemical flux) and purely chemical obstacles (only interfering with chemical flux) elicit similar dynamical behaviors; c) a minimal program for robustly gathering simulated cells does not involve mechanisms for obstacle sensing and avoidance; d) fluctuations of the dynamics concur in preventing multiple stable clusters. Comparing those findings with previous results, we speculate about the fact that chemotactic cells can avoid obstacles by simply following the altered chemical gradient. Social interactions are sufficient to guarantee the aggregation of the whole colony past numerous obstacles.
\end{abstract}


\end{frontmatter}


\section{Introduction}
\label{Introduction}


The social amoeba \textit{Dictyostelium dicoideum}, a model organism in biomedical research \citep{williams2010dictyostelium,strmecki2005developmental}, is a well-studied example of chemotactic life-form \citep{kessin2,futrelle,urushi,bonner,bonner2,jowhar}. Chemotaxis involves the detection of local gradients of chemical signals (spatial detection), the polarization of the cell and the subsequent movement of the cell up the gradient \citep{romeralo, devreotes}. In the absence of nutrition sources, starving cells of \textit{D. discoideum} are able to send and to process periodic stimuli of 3’,5’-cyclic adenosine monophosphate (cAMP) that act as chemoattractant \citep{alcantara,van,nan,robertson}. As a consequence, a scattered colony is able to self-coordinate towards an aggregate bulk \citep{kessin,fey}.\\ 
Since directed migration is a common feature in many cell systems, \textit{D. discoideum} is widely studied \citep{munoz} and has inspired problems of decentralized gathering \citep{bonabeau,vlass}. Consequently, over the years, many aspects of chemotactic behavior in amoebozoan colonies have been investigated both from an experimental \citep{robertson,vicker,darmon,willard} and a theoretical \citep{maree,dallon,dallon2,palsson,nagano,nagano2,pineda,savill} point of view. Remarkably, the colony was shown to perform decentralized gathering in a self-coordinated manner, the aggregation mechanism being an emergent property arising from local rules. However, past studies are mostly considering colonies in a homogeneous environment. On the contrary, \textit{in vivo} conditions are often characterized by heterogeneous environments. Only a few studies have already tackled such issue \citep{elliott,grima} considering the mechanical and rheological properties of the microbiological system at the scale of single cell. On the other hand, attempts to investigate the behavior of a whole colony in a heterogeneous environment rely on Cellular Automata simulations \citep{fates,hatzikirou}. Cellular Automata (CA) provide individual-based models at a population level, but their level of complexity is often not sufficient to grasp important details of a process and can be further improved. A natural extension of CA towards more flexible and realistic models is the Multi Agent Systems (MAS) computational approach. An \textit{agent} is an individual computer system characterized by an arbitrarily complex architecture for behaviors and is capable of independent action on behalf of its user or owner \citep{woolridge, weiss}. In addition to individual realistic representation, additional features of MAS models include embedding of agents into dynamic environment and intracellular decision processes triggered by biochemical cell-cell or cell-matrix interactions \citep{galle}. Examples of MAS models addressing \textit{D. discoideum} gathering in homogeneous environments can be found in \citep{parhizkar, Gallo}.\\

\subsection{Heterogeneous environments}
\label{Methods}
When considering cell colonies, a homogeneous environment is usually modeled as an Euclidean plane ($\mathrm{Env} \in \mathbb{R}^2$) as suggested by \cite{bakerpdes}. Adding obstacles corresponds to alter the topology of the plane according to their size, shape and position. There are two meaningful types of obstacles that are of biological interest. ``Physical'' obstacles are such that they prevent cells to cross and absorb the cAMP flux; ``Chemical'' obstacles absorb cAMP molecules alone, letting amoebas through. The former represents concrete objects, while the latter can be interpreted as chemical species that degrade the cAMP flux locally.\\ Analytically, physical obstacles are represented as locally reflecting border conditions for Amoeba cells. Let $J_{\mathrm{Am}}(x,t)$ be the component of Amoebas flux that is perpendicular to the obstacle; a ``physical'' obstacle with size $\bar{S}_{obs} = \{\mathrm{length}, \mathrm{height} \}$ is defined as a local reflecting boundary condition in the following sense:
\[
\left\{
\begin{array}{c}
J_{\mathrm{Am}}(\hat{x},t)=0 \\
\frac{\partial J_{\mathrm{Am}}(\hat{x},t)}{\partial n_m} \equiv \mathbf{n_m} \cdot \nabla J_{\mathrm{Am}} = 0 \\
\end{array}
\right.
\]
where $\hat{x}$ is the set of coordinates of points belonging to the obstacle edge that is perpendicular to the normal vector of the flux: $\hat{x}_x = \{\mathbf{x'} = \{x', y'\} \; s.t. \; \mathbf{x'} \bot \mathbf{n_m} \}$ with $\mathbf{n_m}$ being the flux normal vector to the m-th edge of the obstacle.\\
From the perspective of the flux of Amoebas, ``chemical'' obstacles do not represent any source of topological alteration as they let agents through.\\ When considering the chemical flux of cAMP molecules, both ``physical'' and ``chemical'' obstacles act as local absorbing boundary conditions:

\[
\left\{
\begin{array}{c}
J_{\mathrm{cAMP}}(\hat{x},t)=0 \\
\frac{\partial J_{\mathrm{cAMP}}(\hat{x},t)}{\partial n_m} \equiv \mathbf{n_m} \cdot \nabla J_{\mathrm{cAMP}} < 0 \\
\end{array}
\right.
\]

\noindent Usually, environmental heterogeneity occurs on a physical scale that is comparable to that of individual cells \citep{grima}. As a consequence, macroscopic continuum models (PDEs) of cell movement have difficulties in solving the problem of heterogeneous domains \citep{bakerpdes}.\\
On the other hand, analytical individual-based models that focus on single cells neglect the complexity of the system: they effectively inquire what happens to the cell movement in response to chemical gradients $\nabla C = g \, \hat{z}$ ($\hat{z}$ being the direction unit vector) around an obstacle, but they do not consider the dynamical interactions among multiple cells behaving as a complex network \citep{boccaletti,saetzler, grune}. As a matter of fact, \textit{D. discoideum} colonies consist of a population of simple agents interacting locally with one another and with their environment. Although there is no centralized control structure, local interactions between agents are capable of driving the emergence of global complex behaviors such as multi-stability, collective choices and self-organization \citep{beni,oconnor,garnier}. \\

\subsection{Robustness for biological systems}

When considering complex biological systems, the notion of \textit{robustness} is crucial to assess their dynamical properties quantitatively. Robustness is broadly defined as the ability of a system to maintain its function despite the presence of perturbations \citep{kitano2004biological}. A way to interpret biological systems is to consider them as computing machines. To their utmost simplification, computing machines deliver an output given an environmental input and a finite set of internal rules (individual program). Computing systems are composed by interacting computing machines. As this notion is congruent to that of MAS \citep{weiss}, said simulation approach is a natural modeling candidate. Within this framework, robustness can be tested as the ability of a computing system to maintain its function despite alterations of individual programs. This way, we can try to evaluate the minimal program (at individual level) that is sufficient to maintain the emerging function of the whole system (at system level). A minimal program $P^{m}$ among a set of possible programs $\mathcal{P} = \{ P \}$, each with length $len(P)$ corresponding to the number of rules, is formalized as:

\begin{equation}
P^m = \argmin_{len(P)} \mathcal{P}
\end{equation}

\noindent In the present context, $\mathcal{P}$ is the set of programs composed by biologically validated rules (e.g. cAMP relay, nutrition sensing \dots) plus additional hypothetical mechanisms (e.g. obstacle sensing). The system function corresponds to its ability to achieve decentralized gathering within time scales that are biologically consistent (e.g. as reported by \cite{fey}). We then conjecture that social interactions among cells are sufficient to ensure robust behavior towards coordinated cell movement and obstacle avoidance, without additional mechanisms for obstacle detection and avoidance.

\subsection{Aim and overview}
The present work aims at investigating the aggregation of chemotactic colonies in heterogeneous environments with obstacles. In particular, as cellular mechanisms for obstacle sensing and avoidance are still poorly understood \citep{grima}, testing concurrent hypothesis in a computational testbed will provide insights on the main processes involved. Concurrent hypothesis are modeled as different programs $P_i \in \mathcal{P}$. By using a Multi-agent system model, we study the efficiency of chemotaxis in achieving controlled and robust cell migration, focusing on dynamical features and self-organization at a population level \citep{camazine}.\\ To complete the investigation on how obstacle affect the dynamics, both ``physical'' and ``chemical'' obstacles are considered and their impact assessed.  \\

\noindent The paper is structured as follows. First, we defend the model choice and we describe the implementation of the MAS model and its environment. Second, we define a topology-independent quantitative measurement to follow the complex evolution of the colony. Third, we present the results: how obstacles (and their type) affect the dynamics and what are the minimal behaviors for cells towards efficient decentralized gathering. Finally, as we notice that obstacles elicit multiple stable clusters of cells, we explore the effect of individual fluctuations on the formation of said clusters.\\ We remark that the present modeling approach is effective both in single-obstacle and multi-obstacle settings and is easily biologically interpretable. Hence, we believe helping researchers to explore additional realistic configurations and test system robustness efficiently.

\section{Methods}
\subsection{The Multi Agent systems model}

In order to perform simulations and analysis, we extend an existing MAS-based computational framework that was purposefully designed to address \textit{D. discoideum} behavior and that has already been tested and validated \citep{Gallo}. The model simulates the desired dynamics after stating individual behavioral rules and setting biologically consistent parameters. Four main agents compose the model architecture: Environment $\mathrm{Env}$ (a squared closed $\mathrm{dim} \times \mathrm{dim}$ domain, divided into cells with associated food sources $b(t)$ possibly growing over time), Amoeba agents $\mathrm{Am}$ (proactive agents representing individual cells), cAMP signals $\mathrm{cAMP}$ (vectorial packages of chemical signal) and Obstacles $\mathrm{Obs}$.\\

\begin{figure*}[h!]
	\centering
	\includegraphics[scale=0.4]{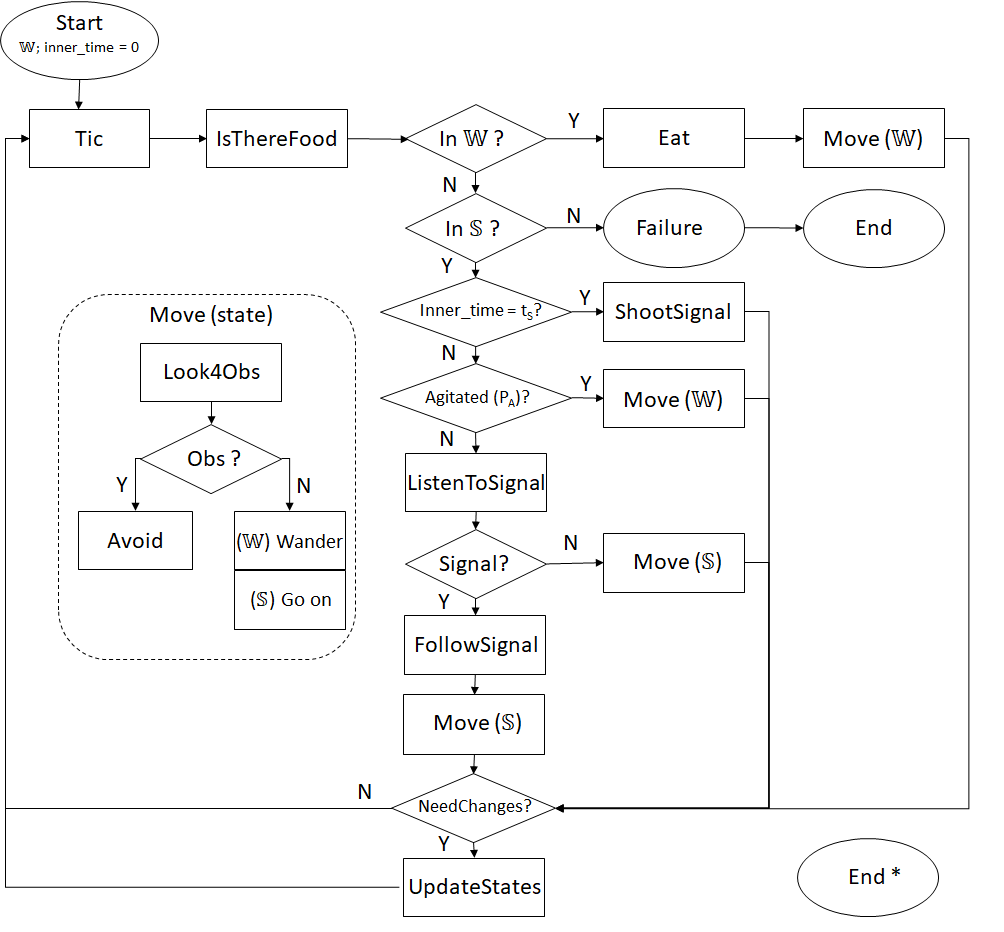}
	\caption{Schematic representation of the decision flowchart for a single agent Amoeba as extended from the MAS model described in \cite{Gallo}. In particular, the \textit{Move} statement can be updated with rules for obstacle detection and avoidance. This program represents a concurrent hypothesis for a robust minimal model.}
	\label{architecture}
\end{figure*}

\noindent Agents Amoeba are implemented with individual behaviors embedded in a complex  architecture: given the set of configurations \{E$_{g}\}$ of the environment $\mathrm{Env}$, the agent is a tuple $\mathrm{Am} = \langle \text{\textit{Inspect, Internal, Action}} \rangle$. \textit{Internal} = \{\textit{Update, States}\} is a set of features  that are not directly interfacing with the environment. They distinguish the agent from a purely reactive one. In particular, \textit{States} = \{S$_{j}$\} is the set that defines the internal state of each agent and that selects other rules accordingly. In the present model, only two states are available: wandering $\mathbb{W}$ (there is food, so the amoeba is eating) and starving $\mathbb{S}$ (the amoeba is \textit{not} finding food, so it begins the aggregation). \textit{Inspect} = \{I$_{i}^{j}$: E$_{g} \mapsto \mathrm{per}$\} looks at E$_{g}$ (locally) and maps it into internal perception \textit{per}. \textit{Update} = \{U$_{l}^{j}$: $\mathrm{per} \mapsto S_{j}$\} may update the states according to the perceptions. \textit{Action} = \{A$_{k}^{j}$: S$_{j} \mapsto E_{g'}$\} is the set of feasible actions on $\mathrm{Env}$ according to the current state of the agent. In Fig. \ref{architecture} the decision-making flowchart for a single agent is reported.\\
Values for variables and parameters of the modek come from settings described in \cite{Gallo} and are derived from validated biological literature. See Table \ref{thr} for simulation values of density $\rho$ and number of agents $N$, for agents' speed $v_A$, cAMP diffusion drift $v_c$, cAMP shooting period $t_s$ and internal ``agitation'' noise $P_A$. The latter is the probability that an individual mis-processes the chemical signal due to asymmetric binding receptor occupancy \citep{van}; it is inserted to improve realism of simulations as stochastic perturbations are often present in biological systems.  This way, stochastic perturbations and individual failures can be easily included and controlled by tuning the respective parameters.\\ 

\begin{table}[h!]
		\centering
		\caption{Meaningful parameters and variables for simulations (set to be biologically consistent). Values from \cite{Gallo}.}
		\label{thr}
		\begin{tabular}{llllll}
			\toprule 
			$\rho [\frac{\mathrm{cell}}{mm^2}]$      &   $N$  &   $v_{A} [\frac{\mathrm{unit}}{s}]$     & $v_{c} [\frac{\mathrm{unit}}{s}]$ & $t_S$ [s]& $P_A$  \\ 
			\midrule
			563 & 2640 & 1.4 & 4.7 & 10 & 0.001   \\
			\bottomrule
		\end{tabular}
	\end{table}
	
\noindent We mentioned that chemical signals are modeled as discrete cAMP packages (tokens). It has been shown \citep{Gallo} that the vectorial message sharing approach generates results that are consistent to those obtained with the typical diffusion of chemicals. In fact, we can relate a chemical gradient $\nabla C = g \, \hat{z}$ with a probability gradient. In this case, of the probability that a starving amoeba processes the information carried by an absorbed cAMP token $P^{amoeba}(\{I_{cAMP}^{starving}: E_{g} \mapsto \mathrm{per} \})$. At the same time, such strategy is cheaper than that of diffusive chemicals in terms of computational cost \citep{nemec,wang2010}, thus allowing better scaling for large populations.\\

\noindent Heterogeneity is modeled by inserting objects $\mathrm{Obs}$ in the environment. In the present work, obstacles are rectangular with arbitrary length and height. As described above, obstacles can be either ``physical'' or ``chemical''. Following the simulation framework, they are modeled with local rules, depending on their kind $k_{obs}$. Both obstacles act as sink for cAMP molecules:\\

\begin{algorithm}[]
\SetAlgoLined

  \If{cAMP overlaps obstacle}{
   delete cAMP token\;
}{}

 \caption{Absorb cAMP}
\end{algorithm}
\vspace{\baselineskip}

\noindent Moreover, Amoeba agents can not cross ``physical'' obstacles, while no restriction is given on ``chemical'' ones:\\

\begin{algorithm}[]
\SetAlgoLined
$v_{A}^{\bot} = \mathbf{v_a} \cdot \mathbf{n_m} $\; 
  \If{$k_{obs}==$ phys}{
  \If{Amoeba overlaps obstacle}{
   $v_{A}^{\bot} = -v_{A}^{\bot} $\;
}
}

 \caption{Reflect Amoeba}
\end{algorithm}
\vspace{\baselineskip}

\begin{figure}[h!]
	\centering
	\includegraphics[scale=0.4]{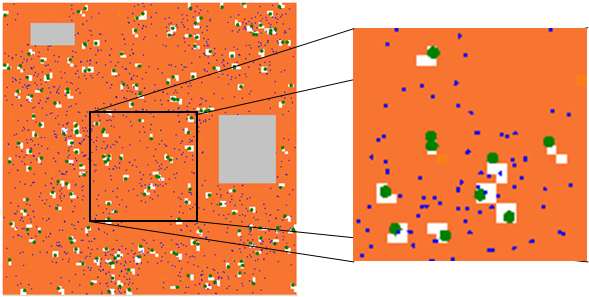}
	\caption{Representation of a typical setting for heterogeneous environment: a squared domain with food sources (orange) contains agents Amoeba (green) that are able to reproduce the sending-sensing-orienting-moving behavior that characterize chemotactical motion in \textit{D. discoideum}. Blue dots represent cAMP tokens, while gray objects are obstacles; white spaces indicate depleted food sources. Original figure colored.}
		\label{env}
\end{figure}

\noindent Fig. \ref{env} shows an example setting of heterogeneous environment with embedded agents (green circles), cAMP tokens (blue dots) and obstacles (grey). Food sources are color coded and can evolve during simulation cycles. In the figure, orange is for $b(t)=1$, white for $b(t)=0$.

\noindent The model focuses on the first aggregating stages of the gathering process (``streaming'' process \citep{romeralo}) and it does not implement adhesion forces between neighbor cells, thus relying on social mechanisms to maintain cell-cell cohesion. Nonetheless, we believe that it represents an appropriate tool to investigate the dynamical properties of complex colonies while coping with topological perturbations due to obstacles. The code is freely available from the GitHub repository \url{https://github.com/daniele-proverbio/amoeba}.

\subsection{Defining quantitative measurements}
Traditional metrics to quantify the aggregation of a scattered colony are: ``Box count'', namely the dimension of a box containing all cells \citep{fates}, and the marginal Variance \citep{Gallo}, accounting for the average spread of the colony around the central bulk. However, such measurements are solely valid in homogeneous environments, as they are susceptible to topological characteristics and fail to track aggregation in multiple clusters.\\
In order to quantify the aggregation stages in heterogeneous environments, we then suggest a ``mean local gathering factor'' metric. In particular, the desired metric needs to be monotonically increasing in time if the colony is steadily aggregating and it must not be directly dependent on the environmental topology. The metric is constructed as follows:

\begin{enumerate}
	\item We define a certain range (in turn defining a ``neighborhood area'') under which two cells are said to be neighbors. The reason for defining a ``range'' is that, as adhesion cells are not implemented in the model, cells are kept close by social interactions through message sharing \citep{Gallo}. As a consequence, two neighbor cells can vary their distance in time by oscillating around each other. For each agent, said `` neighborhood range'' $R_N \in \mathbb{R}$ is defined as follows:
	\begin{equation*}
	R_N = R +v_{A} \cdot \Delta t - r
	\end{equation*}
	where $R$ is the agent radius, $v_A$ is the agent speed, $\Delta t$ is equal to 1 cycle; $r$ is the mean distance from the centers of two overlapping cells, that can be determined by studying preliminary simulations. In the present setting, $r=0.4$ 
	(see Table \ref{thr} for simulation values).
	
	\item In the position space, the ``neighborhood area'' of the $i$-th cell is given by:
	\begin{equation}
	A^*_i = \{ (x',y') \; \mathrm{s.t.} \; (x_i - x')^2 + (y_i - y')^2 \leq R_N^2 \}.
	\end{equation}
	where $(x_i, y_i)$ are the cell center coordinates. At the same time, the physical area occupied by the $j$-th agent is given by:
	\begin{equation}
	A_j = \{(x'',y'') \; \mathrm{s.t.} \; (x_j - x'')^2 + (y_j - y')^2 \leq R^2 \}.
	\end{equation}
	
	\item The set $In(t)_i$ of $j$-amoebas that are neighbors (at time $t$) of the $i$-th agent is the set of $j$-agents whose physical area intersects the $i$-th ``neighbor area'' at time $t$, that is:
	\begin{equation}
	In_i(t) = \{\{\mathrm{Am}_j \}, \, j=1 .. n \; \mathrm{s.t.} \; [\mathrm{Am}_j \cap \mathrm{Am}^*_i](t) \neq \emptyset\}
	\end{equation}
	
	\item The number of neighbor cells for the $i$-th amoeba is thus given by:
	\begin{equation}
	N^{in}_i(t) = \mathrm{dim}\left( In_i(t) \right)
	\end{equation}
	where $\mathrm{dim}$ represents the dimension of the set. $N^{in}(t)$ is the \textit{local gathering factor} and increases the deeper the $i$-th amoeba is in the gathering pattern and/or in the final aggregate.
	
	\item To obtain a \textit{mean local gathering factor}, we average $N_{in}(t)_i$ over the number of amoebas $N$:
	\begin{equation}
	\aleph(t) = \frac{1}{N}\sum_{i=1}^{N} N^{in}_i(t)
	\end{equation}
\end{enumerate} 

\begin{figure}[h!]
	\centering  
	\includegraphics[scale=0.5]{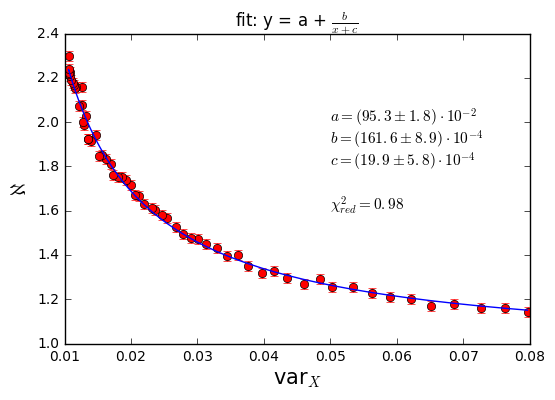}
	\caption{Mean local gathering factor $\aleph$ as defined in this section against marginal Variance as used by \cite{Gallo}. Data come from repeated simulations in homogeneous environments. As expected, when obstacles are absent, the two metrics are inverse one of the other, as guaranteed by the hyperbolic fit (values reported).}
	\label{aleph_vs_var}
\end{figure}

\noindent Note that $\aleph$ will not be normalized, since the maximum value it can reach at the end of the simulation time $t_{max}$ is itself informative about the ``strength'' of the gathering process.\\
In a homogeneous environment, mean local gathering factor $\aleph$ is precisely the inverse of the marginal Variance $var_X$, as the former quantifies how much agents are packed together while the latter measures how spread the colony is. Fig. \ref{aleph_vs_var} shows the agreement between data (from repeated simulations) and the following fit relationship:
\begin{equation}
    \aleph = a + \frac{b}{var_X + c}
\end{equation}

\subsection{Simulation protocol and remarks}

\begin{figure*}[h]
	\centering
	\includegraphics[scale=0.5]{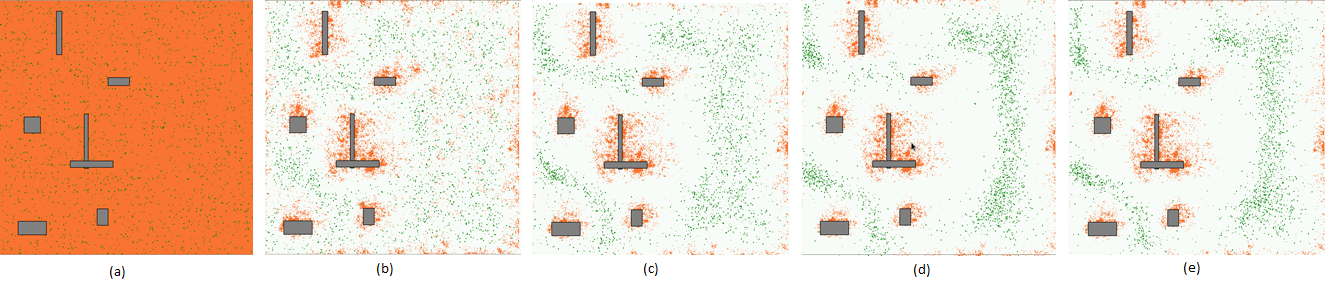}
	\caption{Evolution of the colony from animation. Measurements for the simulation are shown in Fig. \ref{aleph1}. (a) Setting the simulation with constant food sources in $\mathrm{Env}$; (b) t =400, when measurement starts after the neglected transient; (c) t = 700, we observe different areas with different cell density; (d) t = 1000: multiple sub-domains with local dense clusters are recognizable (left-hand side). The obstacle-free half domain (right-hand) is characterized by the ``usual'' aggregation pattern: here, the colony is streaming and gathering as in previous studies \citep{dallon, palsson, Gallo}, shaped by the rectangular domain. (e) t = 1300 the metastable clusters are even more recognizable. Note that, during the evolution, most food sources have been eaten, meaning that Amoeba cells explored most of the domain apart from regions close to the border and the obstacles. Original figure colored.}
	\label{subs_process}
\end{figure*}

Simulations and subsequent analysis are based on the following protocol.
\begin{enumerate}
	\item Consider the parameter space $\mathbb{PS}$ defined by the number, size, location and kind of obstacles: $\mathbb{PS} = \{n_{obs} \times \hat{\mathrm{S}}_{obs} \times \bar{x}_0 \in \mathrm{Env} \times k_{obs} \}$. Remember that they are initially considered of rectangular shape. The first aim is not to completely investigate $\mathbb{PS}$, but to verify whether a certain configuration elicits local adaptation of the dynamics. Therefore, initial settings are chosen so to get two distinct sub-domains of the original $\mathrm{Env}$, respectively with and without obstacles. This way we can compare what happens when obstacles are placed locally and whether the obstacle-free sub-domain elicits the same aggregation pattern as when $n_{obs} = 0$. Consequently, simulations are run by considering random settings from $\mathbb{PS}$ and the sub-domains constraint.
	\item Set external noise sources to zero, since the main focus is on perturbation caused by obstacles.
	\item Run repeated simulations of early aggregation in order to increase statistic relevance, thus measuring ($\aleph(t) \pm \sigma_{\aleph}(t))$. For simulation parameters and variables, refer to Table \ref{thr}. 
	\item As the simulations described in \cite{Gallo} are specific for the pre-aggregation stages of the gathering process (``streaming''), we remain consistent in considering the same time interval for the analysis. In simulation cycle units, time limits are $400 < t < 1300$; this way we neglect the uninformative initial transient and stop the simulation when the aggregate starts to appear, thus focusing on the ``streaming'' process.
	\item Study the influence of $k_{obs}$.
	\item Contrast different programs $P_i \in \mathcal{P}$ to search for a minimal program $P^m$ that ensures robust aggregation in heterogeneous environments.
\end{enumerate}

\section{Obstacles as local source of perturbation}
\label{sec:3}

\begin{figure}[h!]
	\centering  
	\includegraphics[scale=0.5]{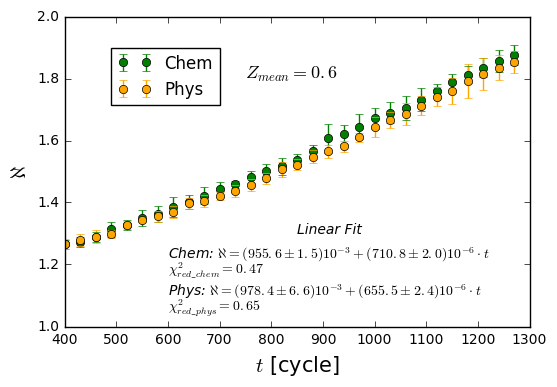}
	\caption{Evolution of the colony during early aggregation time interval (as described in the text). More than 30 repeated simulations were performed with either physical and chemical obstacles. ($\aleph \pm \sigma_{aleph}$) is monotonically increasing in time, as we would expect in case of a steadily aggregating colony (fit curves are not shown for clarity reasons). Simulations with different obstacle configurations $\in \mathbb{PS}$ gave the very same qualitative result.
		Moreover, a $Z$ test confirms that the experiments with different $k_{obs}$ are not statistically separate. $Z_{mean} = 1/N \sum_{i=1}^N Z_i$ summarises the average value for the $Z$ test at each i-th time point. Original figure colored.}
	\label{aleph1}
\end{figure}

By following the above protocol, we first inquire whether obstacles indeed represent local sources of perturbation. We run repeated simulations of the same setting, which was randomly chosen from $\mathbb{PS}$, measuring $(\aleph \pm \sigma_{\aleph})$ and also considering simulated animations. $k_{obs}$ was varied from ``physical'' to ``chemical''.\\
As we see in Fig. \ref{subs_process}, obstacles perturb the colony dynamics locally. On the left-hand half of the domain, we observe that obstacles circumscribe sub-domains with different cell densities, thus eliciting metastable clusters. On the contrary, in the right-hand side (obstacle-free region) we recognise the usual streaming behavior towards the aggregate, as reported in previous studies \citep{dallon,palsson,Gallo}. Moreover, as shown in Fig. \ref{aleph1}, the measure $\aleph$ is monotonically increasing, as we would expect in case of a steadily aggregating colony. This fact quantifies that clusters have been formed despite the presence of obstacles, which do not cause the cells to remain scattered but prevent the formation of a global aggregate.\\

\noindent Animations are also informative when we consider the distribution of food sources (orange areas). Food is initially set constant ($b(0) \in \mathrm{Env} = 1$, w.l.o.g.) and is not programmed to augment over time.
    \[
    b(t+1)-b(t)=
    \begin{cases}
    -1, & \text{if $\mathrm{Am}$ is eating,} \\
    0, & \text{otherwise}
    \end{cases}
    \]
Hence, during the simulation, white space indicates regions that have been explored by amoebas, while orange ones have not. In Fig. \ref{subs_process} we observe that $\mathrm{Am}$ agents explored most of the domain during the aggregation, but they seldom went close to obstacles. In addition, we observe that physical and chemical obstacles elicit aggregating behaviors that are not statistically distinguishable (see Fig. \ref{aleph1}). On the one hand, it suggests that, colony-wise, chemical and physical obstacles elicit the same confining effects by disrupting the flow of cAMP chemicals that guide the cells. On the other hand, this evidence asks for a more detailed investigation of the individual programs (at cellular level) that guide amoebas towards robust gathering.

\section{Emerging behavior for obstacle avoidance}

Given the previous findings regarding the impact of obstacles on the gathering process, we now ask what is the minimal program $P^m$ that guarantees a robust aggregation process. We recall that performing MAS simulations corresponds to check for the emergence of complex phenomena (coordinated aggregation past obstacles) from microscopic rules (individual programs for Amoeba agents). Hence, $P^m$ refers to a program at cell level leading to robust aggregation of the whole colony (population scale). Here, \textit{robust} only refers to the fact that cells form clusters without being scattered all over the environment. Whether clusters are unique or manifold is not currently inquired. \\
To do so, we update the \textit{Move} rule for amoeba agents (see Fig. \ref{architecture}). This results in having two concurrent programs for the agent Amoeba. $P1$ corresponds to the original set of rules for $\mathrm{Am}$: an agent's movement is always up gradient. $P2$ corresponds to the updated \textit{Move} statement: an agent first checks for nearby obstacles and avoids them. In pseudocode, $P1$ and $P2$ read as follows:

\begin{lstlisting}[language=Python]
def P1
    inherit from 
        Am = {Inspect, Internal, Action}
end
\end{lstlisting}

\begin{lstlisting}[language=Python]
def P2
    inherit from 
        Am = {Inspect, Internal, Action}
    update Move:
        isObs = checkFor(nearbyObstacle)
        if isObs == true:
            avoidObstacle
end

\end{lstlisting}

\begin{figure}[h!]
	\centering  
	\includegraphics[scale=0.5]{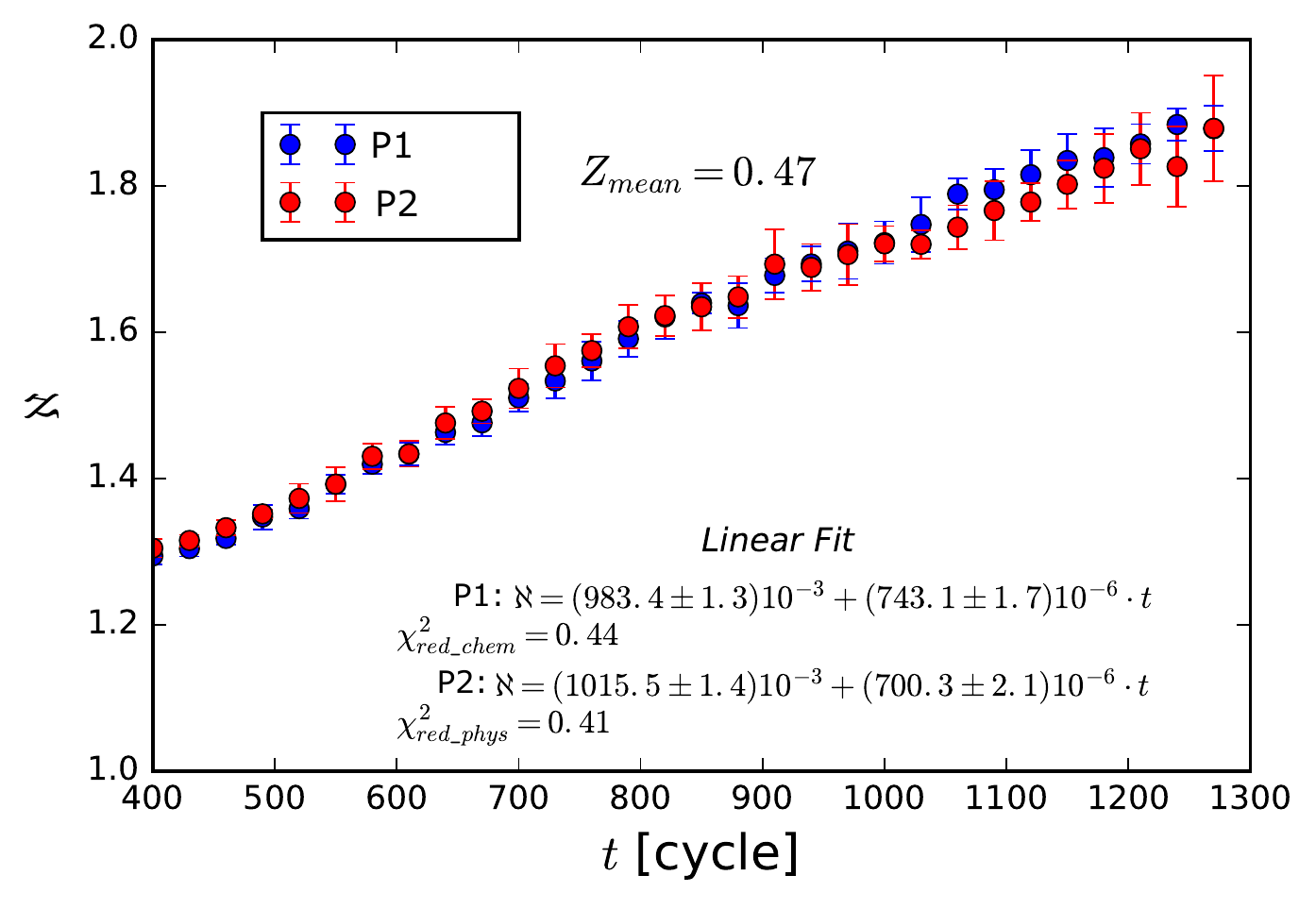}
	\caption{Evolution of $(\aleph \pm \sigma_{\aleph})$ during the colony aggregation. Two experiments for $P1$ and $P2$ are contrasted (obstacle settings are kept constant). Although not shown for clarity reasons, a linear fit was performed to confirm the increasing trend; $\chi^{2}_{red}$ values are shown to estimate the goodness of fit. A test $Z$ was also performed and its mean value reported in the chart. It shows that dynamic behaviors with or without updated \textit{Move} code are not statistically separate.}
	\label{aleph2}
\end{figure}

\begin{figure*}[h!]
	\centering
	\includegraphics[scale=0.5]{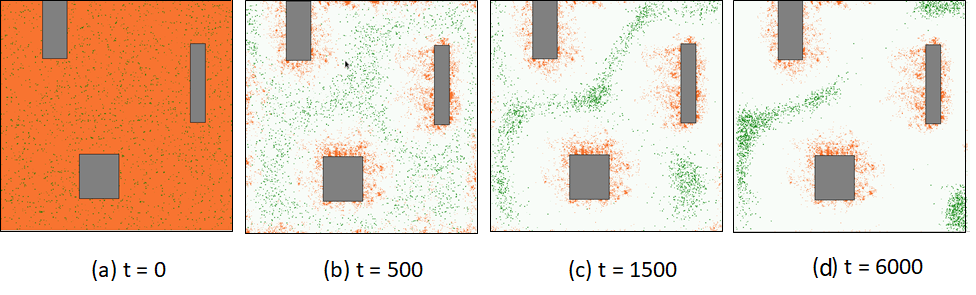}
	\caption{Several snapshots during the evolution of a colony. They refer to the experimental set whose measurements are reported in Fig. \ref{aleph1}. In order to magnify the spontaneous formation of multiple clusters, simulations were let run after the streaming interval considered in the above analysis ($400<t<1300$). Obstacles are randomly scattered in the whole domain. (a) Setting the experiment with constant food sources in $\mathrm{Env}$; (b) $t=500$: after an initial transient, streams appear; (c) At $t=1500$ early aggregation after streaming is almost completed. (d) $t> 1500$: Multiple clusters are clearly recognizable: there are three main attractors around which amoebas are gathering. Note that cells have explored almost all the space (they ate all bacteria that were deployed initially.)}
	\label{3obs}
\end{figure*}

\noindent Note that the two separated programs are dual to Algorithm 2, apart from the fact that are cell-oriented. Implementation is reported in the publicly available code.\\
In this case, we are not interested in maintaining two sub-domains as in Sec. \ref{sec:3}, so obstacles with different shape and size are randomly scattered all over the environment. To test for mechanisms for obstacle sensing and avoidance, obstacles are always of ``physical'' kind. Repeated MAS simulations are then performed, testing both programs in different settings. Fig. \ref{3obs} shows the evolution of a simulation without updated \textit{Move}: we observe the ability of the colony to overcome obstacles and to stream toward multiple clusters even in the absence of specific mechanisms. In fact, similarly to what reported in Fig. \ref{subs_process}, cells follow chemical gradients away from obstacles. More quantitatively, we measured $\aleph(t)$ for repeated simulations in different settings, while changing program. Results are reported in Fig. \ref{aleph2}. We observe that evolution of the colony whose individuals follow $P^1$ is not statistically distinguishable from that ruled by $P^2$. This evidence is strengthened by a $Z$ test. Other considered settings in the parameter space $\mathbb{PS}$ give the same result. 

\noindent Previous works \citep{grima,elliott} suggested that cells are not required to own specific sensing mechanisms, as they can in many cases avoid the obstacle by following the perturbed chemical gradient in its vicinity. Such hypothesis was preliminary tested with single-cell individual-based models. Using a multi-cell scheme, we suggest that similar results can be obtained when considering the whole complex system: at a population level, cells do not even came close to obstacles as they are guided by social interactions through message sharing. From the ``perspective of the colony'', its microscopic components (individual amoeba cells) are sufficiently equipped with $P1 = P^m$ to get the emergence (at population scale) of coordinated aggregation past an obstacle.\\
A speculative biological interpretation is the following: in order for a colony to robustly aggregate, it is not necessary for a cell to develop specific mechanisms for obstacle avoidance. Cells seem to be guided by chemicals that, once being absorbed by obstacles, highlight the best paths for chemotaxis-directed migration. The self-aggregating patterns of the colony as a whole not only guide single cells towards stable clusters, but they also elicit collective choices to avoid obstacles.

\section{Fluctuations drive the system out of local minima}

\begin{figure*}[h!]
	\centering
	\includegraphics[scale=0.4]{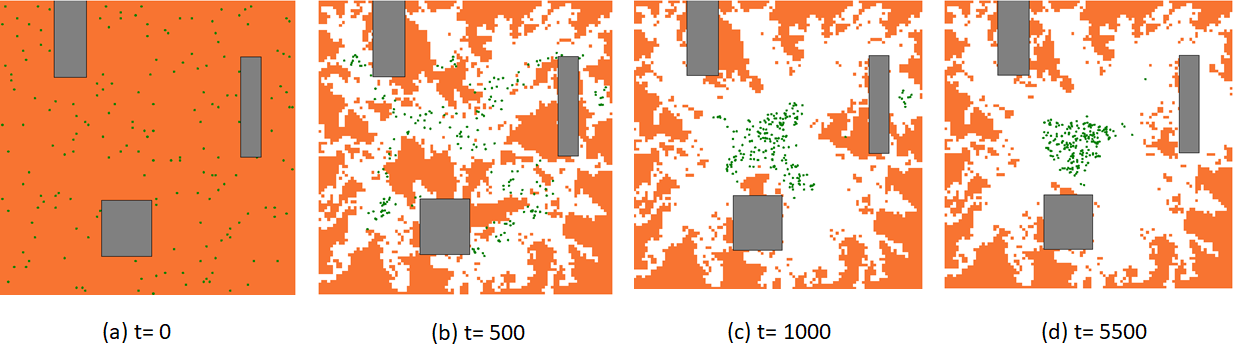}
	\caption{Several snapshots during the evolution of an ``off-scale'' colony, set as described in the text. In order to magnify the final formation of clusters, simulations were let run after the streaming interval considered in the above analysis ($400<t<1300$). For better comparison, obstacles are scattered in the domain consistently with ``on-scale'' simulations (compare with Fig. \ref{3obs}). (a) Setting the experiment with constant food sources in $\mathrm{Env}$; (b) $t=500$: after an initial transient, streams appear; (c) At $t=1500$ early aggregation after streaming is almost completed past obstacles. (d) $t> 1500$: a single cluster is formed, as cells managed to go past obstacles in a coordinated manner, driven by individual fluctuations that prevent local attractors.}
	\label{3obs_off}
\end{figure*}

As described in the above section, cells tend to form multiple clusters when obstacles are present. Fig. \ref{3obs} shows this phenomenon. Borrowing vocabulary from statistical mechanics, multiple clusters correspond to different attractors, that are local minima of a hypothetical energy landscape for the system. Hence, obstacles contribute to barrier potentials between these minima. It is now interesting to inquire if it is possible to get a single cluster corresponding to a global minimum.\\ 
In statistical mechanics, noise is known to push system states out of local attractors. Also, in simulations of biological systems, fluctuations around a mean behavior are known to elicit better explorations of the space \citep{garnier,eber}. Therefore, MAS simulations are now performed to check if fluctuations alone are sufficient to drive the colony towards a single cluster, and how often. As a preliminary search, we consider fluctuations due to the colony being ``off-scale''.\\

\noindent It is known that, for a MAS simulation to be repeatable and stable, number $N$ and density $\rho$ of agents have to considered \citep{Gallo}. Initial values listed in Table \ref{thr} respect the conditions for the system to be considered in a scale-free regime. On the contrary, ``off-scale'' systems are such that individual fluctuations dominate the dynamics. They can not be considered for systematic studies as above, but they are perfect to test noisy settings in heterogeneous environments.\\
We then set new simulations on off-scale configurations by changing density and numerosity values (for instance, $\rho=\frac{200}{100^2}$, $N=200$). See \cite{Gallo} (Fig. 11) for all values $\{ N, \rho\}$ that define an ``off-scale'' regime. Different settings and different configurations of obstacle number and placement are then tested. Fig. \ref{3obs_off} shows an example simulation in which the ``off-scale'' colony aggregates in a single cluster. Table \ref{table2} summarizes how many times (in percentage, out of $200$ simulations) multiple clusters are formed, depending on the system regime (``on'' or ``off'' scale) and how many obstacles are present in the domain (1 or multiple).

\begin{table}[h!]
	\centering
	\caption{Chances [\%] to get multiple stable clusters (at least 2). Data from multiple (200) simulations with different obstacles configurations.}
	\label{table2}
	\begin{tabular}{lll}
		\toprule 
		&   on-scale  &   off-scale    \\ 
		\midrule
		1 obstacle & 15 & 3   \\
		\midrule
		multiple obs & 70  & 30 \\
		\bottomrule
	\end{tabular}
\end{table}

\noindent Table \ref{table2} reports that noisy systems (``off-scale'') are more likely to converge to single clusters than ``on-scale`` ones, which are less prone to individual fluctuations. Moreover, values in Table \ref{table2} show that a single obstacle produces less multiple clusters than multiple obstacles. This suggests that obstacles act as potential barriers between energy minima, thus stabilizing local attractors. However, in highly heterogeneous environments, fluctuations alone are not sufficient to achieve robust and repeatable decentralized gathering in a single cluster. Collective behavior towards aggregation seems to result from a delicate trade-off between robust chemotaxis and random fluctuations as suggested by previous experimental works \citep{dallon2011understanding}.

\section{Conclusion}
The main aim of this project was to investigate how colonies of chemotactical cells behave in the presence of obstacles, as often happens \textit{in vivo}. To do so, we chose the social amoeba \textit{Dictyostelium discoideum} as model organism and we studied its aggregation process. Since we decided to focus on dynamical features elicited by nonlinear social interactions, rather than on individual mechanical properties, we used a Multi Agent System model that had been purposefully designed, implemented and validated. A suitable metric (the \textit{mean local gathering factor $\aleph$}) was also defined.\\
By analyzing chemical and physical obstacles, we noticed that their main impact on the system is to perturbs the chemical flux. A group of cells in collision with an obstacle can in many cases avoid it by simply following signals coming from other directions. In fact, as far as the present model is concerned, agents Amoeba are able to aggregate robustly both when having or lacking specific rules for obstacle sensing and avoidance. Such evidence suggests that, from a population point of view, social interactions are effective to drive the colony past obstacles. In fact, big colonies are only locally perturbed by the presence of obstacles and emergent dynamics is sufficient for robust decentralized gathering in heterogeneous environments. Additional individual abilities might elicit better efficiency, but such is not the case observed in the present simulations. This may suggest why specialized biological mechanisms for avoiding obstacles are only known for a few cells and organisms \cite{grima}. \\
Moreover, preliminary results suggest that a fluctuating system can better explore the state space and overcome obstacles with less probability of forming multiple clusters. On the other hand, as known in the literature \citep{fates,Gallo}, perturbed colonies are less likely to follow the chemical gradient, so an efficient gathering process should present a trade-off between chemotactical stability and random exploration. This is typically suggested in the field of system control \citep{iglesias2010control}. \\

\noindent Often, organisms that live in colonies have little concern for an individual fate, whereas they are evolutionary competitive as a whole population \cite{bonabeau}. Therefore, focusing on the emergence of collective behaviors from individual programs provide additional information on the development of such colonies. We believe that MAS models, that couple cell and population scale, are useful when addressing complex microbiological systems in heterogeneous environments. Further \textit{in vivo} and \textit{in vitro} studies are recommended to test the hypothesis made in this article and to improve our knowledge on behavior of cellular motility in heterogeneous environments.\

\section*{Code}
The code to perform MAS simulations is freely accessible at the following GitHub repository: \url{https://github.com/daniele-proverbio/amoeba}. Statistical analysis was performed with Python open libraries.

\section*{Acknowledgements}
This research has been possible thanks to the access to the INFN Computing Farm in Turin. D.P. would like to thank dr. J. Markdahl for valuable comments.\\

\section*{Disclosure}
This research did not receive any specific grant from funding agencies in the public, commercial, or not-for-profit sectors.\\

\section*{Author contributions}
D.P and M.M. designed and conceived this project. D.P. performed the experiments and the analysis. D.P. and M.M. interpreted the results and wrote the manuscript. All authors contributed to and approved the manuscript.



\bibliographystyle{apalike}
\bibliography{bibliography.bib}







\end{document}